\documentclass{article}
\usepackage{booktabs} 
\usepackage{subfigure}
\usepackage{makecell}
\usepackage{float}
\usepackage{spconf,amsmath,graphicx,amssymb,subfig}

\begin{document}

\title{VITS-based Singing Voice Conversion System with  DSPGAN post-processing for SVCC2023}
%
\name{Yiquan Zhou$^{1,3}$, Meng Chen$^{2}$, Yi Lei$^{3}$, Jihua Zhu$^{1*}$\thanks{* Corresponding author.}, Weifeng Zhao$^{2}$}
\address{
$^1$School of Software Engineering, Xi'an Jiaotong University, Xi'an, China\\ $^2$Lyra Lab, Tencent Music Entertainment, Shenzhen, China\\
  $^3$Audio, Speech and Language Processing Group (ASLP@NPU), School of Computer Science, \\ Northwestern Polytechnical University, Xi'an, China
} 



\maketitle

\begin{abstract}

This paper presents the T02 team's system for the Singing Voice Conversion Challenge 2023 (SVCC2023). Our system entails a VITS-based SVC model, incorporating three modules: a feature extractor, a voice converter, and a post-processor. Specifically, the feature extractor provides F0 contours and  extracts speaker-independent linguistic content from the input singing voice by leveraging a HuBERT model. The voice converter is employed to recompose the speaker timbre, F0, and linguistic content to generate the waveform of the target speaker. Besides, to further improve the audio quality, a fine-tuned DSPGAN vocoder is introduced to re-synthesise the waveform. Given the limited target speaker data, we utilize a two-stage training strategy to adapt the base model to the target speaker. During model adaptation, several tricks, such as data augmentation and joint training with auxiliary singer data, are involved. Official challenge results show that our system achieves superior performance, especially in the cross-domain task, ranking 1st and 2nd in naturalness and similarity, respectively. Further ablation justifies the effectiveness of our system design.

\end{abstract}
\section{Introduction}
\label{sec:intro}

Singing Voice Conversion (SVC) aims to convert the source singing voice to the voice of the target singer or speaker while preserving of original lyrics and melody. It's more challenging than typical speech voice conversion (VC) due to the richer expressiveness in singing voice~\cite{huang2023singing}. 

The key problem for SVC is to decouple and recompose the speaker timbre and the underlying content and melody in the singing voice. To achieve this goal, adversarial training~\cite{chou2018multitarget} is usually utilized to conduct speaker disentanglement by reducing the correlation between speech components. Following the successful application of generative adversarial network (GAN) in VC scenario~\cite{zhang2020gazev,chen2022efficient}, many studies~\cite{lu2020vawgan,9023162,sisman20_odyssey} attempt to employ GAN to improve the SVC performance. However, these systems usually suffer from a trade-off for naturalness and speaker similarity, since the insufficient disentanglement of the timbre and linguistic content in speech. For more robust performance in SVC, recent advancements have been achieved by employing a recognition-synthesis schema, which adopts pre-trained models to extract decomposed speaker-independent linguistic prior and then perform singing voice conversion by an SVC model. The most popular way~\cite{Tian2020TheN,wang2021enriching,guo2020phonetic} is to leverage phonetic posteriorgram (PPG) or neural bottleneck feature (BNF), computed from an automatic speech recognition (ASR) model, as a speaker-independent intermediate representation. Recently, the self-supervised learning (SSL) models~\cite{hsu2021hubert,baevski2020vqwav2vec} trained on massive unlabeled speech data show impressive performance in the downstream ASR task, which is believed that the SSL model can summarize the robust linguistic-related information from speech utterance. Several studies~\cite{jayashankar2023self,soft-vc-2022} investigate the use of SSL features and show promising performance in terms of naturalness and similarity on SVC tasks. 
Depending on the recognition-synthesis schema, some studies~\cite{liu2021fastsvc,li2022hierarchical} focus on the end-to-end framework to directly generate the waveform of the converted singing voice to achieve high-fidelity conversion.

Despite the considerable progress in SVC, it is hard to fairly compare across different approaches, due to the fact that the datasets and evaluation metrics vary across different systems. As a new edition of the voice conversion challenge series, the Singing Voice Conversion Challenge (SVCC)~\cite{huang2023singing} has been recently launched to address this concern. Consistent with its previous iterations, the primary objective in SVCC is speaker conversion. Providing a common test bed, SVCC has set up two tasks: in-domain and cross-domain. The in-domain task obliges participants to build an any-to-one conversion model utilizing the singing data of the target singer. While the cross-domain task, where only speech data of the target speaker is available, is more challenging.

\begin{figure*}[htbp]
  \centering
\begin{minipage}{0.4\linewidth}
      \subfigure[Training]{
      \includegraphics[width=1\columnwidth]{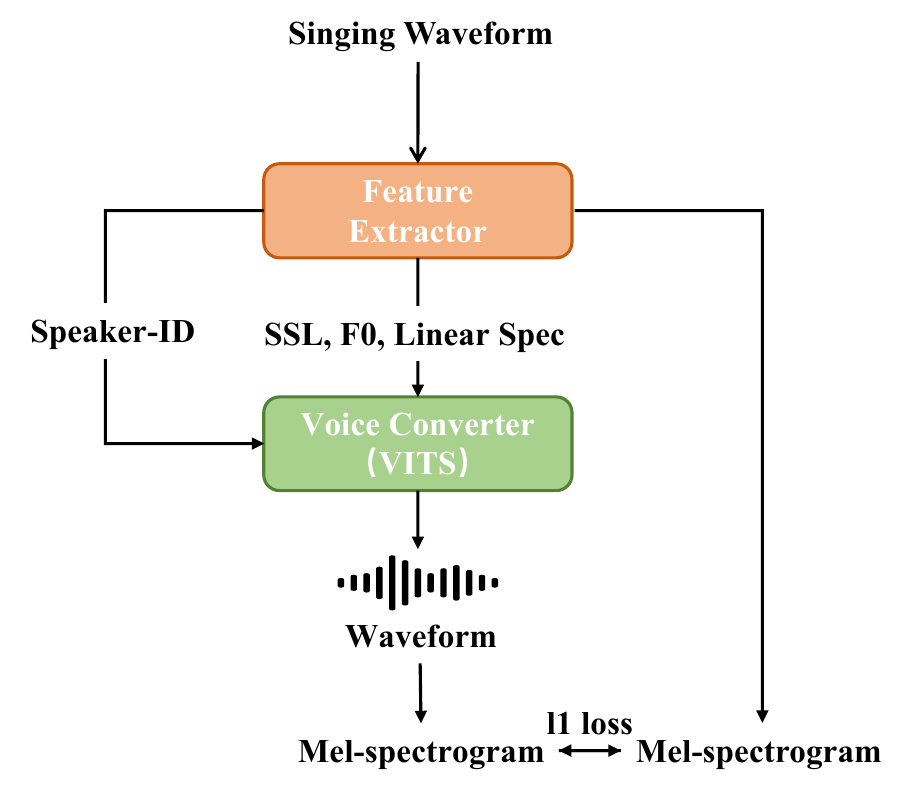}}
\end{minipage}
\begin{minipage}{0.35\linewidth}
       \subfigure[Inference]{
      \includegraphics[width=1\columnwidth]{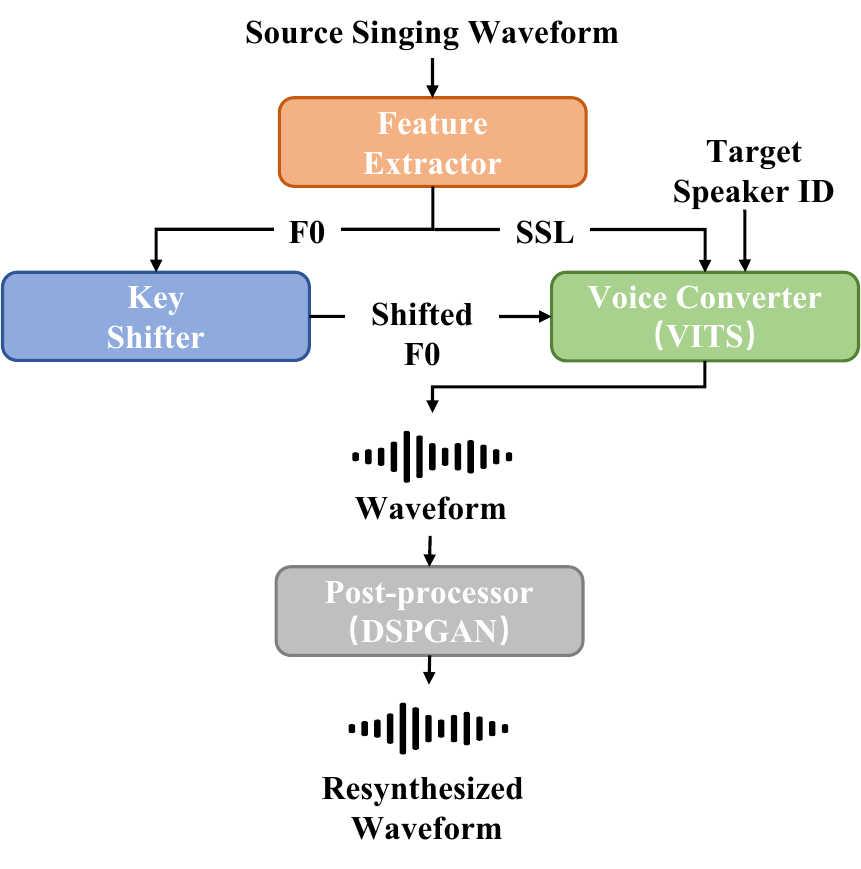}}
\end{minipage}

  \caption{Schematic diagram of the proposed SVC system}
  \label{fig:1}
\end{figure*}

In this paper, we introduce our submitted SVC system, denoted as T02 by the challenge organizer. We implement our SVC model based on the VITS~\cite{kim2021conditional}  architecture. Specifically, following the recognition-synthesis scheme, we employ an SSL model HuBERT~\cite{hsu2021hubert} to extract SSL features as the linguistic content representation of the source singing voice, while fundamental frequency (F0) and speaker ID are used to represent the prosody of the singing voice and the target singer timbre, respectively. Our SVC model fuses the SSL features, F0, and speaker ID to generate the target singing voice. Besides, to alleviate the sound quality artifacts, such as metallic electric sounds, that may exist in the converted results, we further introduce a post-processing step to re-synthesis the singing voice by DSPGAN~\cite{song2023dspgan}, a recently proposed GAN-based universal vocoder for high-fidelity speech synthesis. Additionally, due to the limited amount of target singer data provided by SVCC 2023, our SVC model is first pre-trained in speech and singing data sequentially and then adapted to the target singer with the joint training of auxiliary singer data. Official challenge results show that our system has superior performance in the cross-domain task~\cite{huang2023singing}, ranking 1st and 2nd respectively in naturalness and similarity. The conversion samples of our system can be found on our demo page\footnote{https://zirrtu.github.io/SVCC2023/}.

\section{System overview}
\label{sec:pagestyle}
Fig.~\ref{fig:1} illustrates the schematic diagram of our SVC system, including the training phase and inference phase. The training phase is composed of a feature extractor and a voice converter. For the inference phase, besides the two major modules, a key shifter and a post processor are further adopted.

\textbf{Training:} During the training of the VITS-based voice conversion model, the feature extractor first extracts the SSL feature, F0, and spectrogram from an utterance. Then, the VITS model takes these features as input to reconstruct the utterance with the current speaker ID. Reconstruction loss is adopted during training, which measures the difference between the Mel-spectrogram of synthetic waveform and the Mel-spectrogram of the ground truth waveform.

\textbf{Inference:} During inference, since different singers specialize in different pitch ranges, we adjust the mean value of the F0 sequence according to the different F0 ranges between the source and
target singers by a key shifter. And the voice converter takes
the extracted SSL feature and shifted F0 as input to generate the target singing voice with the target speaker timbre. To further improve the generation quality, an extra post-processor is employed to re-synthesis the converted singing voice.

\begin{figure*}[ht]
\centering
\includegraphics[scale=0.65]{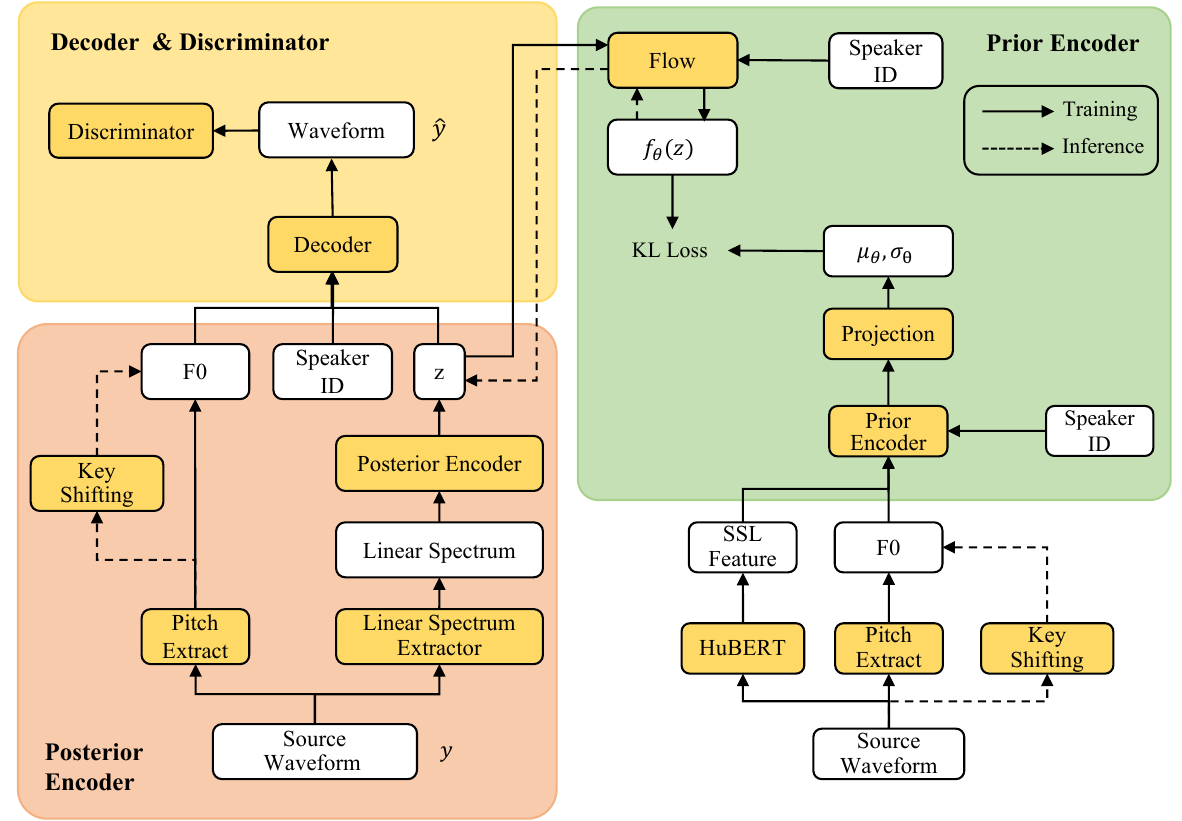}
\caption{The architecture of the VITS-based voice converter.}
\label{fig:2}
\end{figure*}



  %
  %

\section{Modules}
\label{sec:pagestyle}
The design of the feature extractor, the voice converter, the key shifter, and the post-processor are detailed in this section.

\subsection{Feature Extractor}
As mentioned in Section~\ref{sec:intro}, the key factor of SVC is to decouple the speaker timbre and linguistic content in the singing audio. In recent SVC studies~\cite{jayashankar2023self,soft-vc-2022}, features from self-supervised learning (SSL) model, such as HuBERT~\cite{hsu2021hubert}, show robustness in representing linguistic content of the singing voice. Specifically in this paper, we employ a variant version of HuBERT~\cite{pmlr-v162-qian22b} to extract the content representation. For the speaker identity, a look-up table (LuT) is adopted to learn the speaker embedding. Besides the speaker identity and extracted SSL feature, the F0 contour computed by PYIN~\cite{inproceedings} is also adopted in our system to produce an accurate melody of the converted singing voice.

\subsection{Voice Converter}
As shown in Fig.~\ref{fig:2}, we implement our SVC model based on VITS~\cite{kim2021conditional}, which includes four parts: a posterior encoder, a prior encoder, a decoder, and a discriminator. Specifically, during training, the posterior encoder encodes the source waveform $y$ to hidden representation $z$, which models the posterior distribution $P(z|y)$. Then the decoder reconstructs $z$ to the original waveform and forms a self-reconstruction schema with the posterior encoder. Different from the HIFI-GAN decoder in VITS, we extract a sine-based excitation signal~\cite{wang2019neural} from F0 and add it to the hidden feature of the HIFI-GAN decoder~\cite{kong2020hifigan} to improve the singing voice reconstruction quality.
Moreover, a multi-period discriminator (MPD) and a multi-scale discriminator (MSD) are employed to constrain the quality of the waveform in an adversarial manner. The prior encoder fuses singer timbre, pitch, and linguistic content, which models prior distribution. The convertible flow is adopted to transform the prior distribution to the posterior distribution. During the inference, the concatenation of the prior encoder and decoder converts the source singing voice to the target singing voice given the desired speaker timbre.

\subsection{Key shifter}
It's known that different singers specialize in different pitch ranges. Using $F0$ of the source for direct conversion may reduce the target speaker similarity in the converted results. To mitigate this limitation, we incorporate a key shifter during the inference process. First, we pre-calculate the average pitch of the target singer ( $\overline{F0}_t$) and the source ($\overline{F0}_s$). Next we calculate the difference between $\overline{F0}_s$ and $\overline{F0}_t$, resulting in $\delta F0$. Then, $F0$ of the source is added to $\delta F0$ to obtain the shifted pitch $F0_p$. Finally, we use the derived $F0_p$ as input to the voice converter. This simple adjustment effectively improves the speaker similarity in the conversion results.
Note that for the cross-domain task, where only the target speaker's speech is available,  we utilize the average pitch of the target singer in the in-domain task for pitch shifting.

\subsection{Post-processor}
In our SVC system, the conversion model is designed as an end-to-end structure to reconstruct the waveform directly. However, the singing voice generated by the conversion model still exists artifacts such as metallic noise sounds, especially in breathing.

\begin{table*}[]
\centering
  \caption{Speech and singing data for model training.}
\setlength{\tabcolsep}{0.6mm}
\small
\renewcommand\arraystretch{1.9}	
 \begin{tabular}{lcccl}
    \toprule
    Dataset & Duration (hrs) & Type & \hspace*{0.6cm}Usage\\
    \midrule
    \textit{VCTK}~\cite{yamagishi2019vctk}  & 44 & Speech  & \hspace*{0.6cm}VITS Pre-train\\
    \makecell[l]{\textit{Mixed Speech} \\(LJspeech~\cite{ljspeech17}, VCTK~\cite{yamagishi2019vctk}, LibriTTS~\cite{zen2019libritts}, HI-FI TTS~\cite{bakhturina2021hifi})}   & 951.6 & Speech &  \hspace*{0.6cm}DSPGAN Pre-train\\
    \textit{SVCC Cross-domain Task} & 0.18 & Speech & \hspace*{0.6cm}VITS Finetune\\
    \textit{SVCC In-domain Task} & 0.41 & Singing & \hspace*{0.6cm}VITS Finetune \\
    \makecell[l]{\textit{Mixed Singing} \\(NUS48e~\cite{6694316}, Opencop~\cite{wang2022opencpop}, M4singer~\cite{zhang2022msinger}, Opensinger~\cite{huang2021multi})} & 73.3& Singing & \hspace*{0.6cm}\makecell{VITS Pre-train, \\ DSPGAN Finetune }\\
    
    \bottomrule
  
  \end{tabular}
  \label{tab:1}
\end{table*}

To improve the quality of the generated audio, we introduce DSPGAN~\cite{song2023dspgan}, which is a GAN-based universal vocoder for high-fidelity speech synthesis, as a post-processer. In particular, DSPGAN uses sine excitation as the time-domain supervision to improve harmonic modeling and eliminate various artifacts in the GAN-based vocoder, effectively removing the artifacts observed in our converted voice. Moreover, DSPGAN harnesses the mel-spectrogram following extractions from the waveform produced by a DSP module as the time-frequency domain supervision to the GAN-based vocoder. This technique eliminates the mismatch problem caused by the ground-truth spectrograms and the predicted spectrograms from the acoustic model. 
Specifically, we extract the mel-spectrogram of the singing voice obtained from the voice converter and feed it into DSPGAN to acquire the higher-quality waveform.

\begin{figure*}[htb]
\centering
\begin{minipage}{0.4\linewidth}
      \subfigure[English native speakers]{
      \includegraphics[width=1\columnwidth]{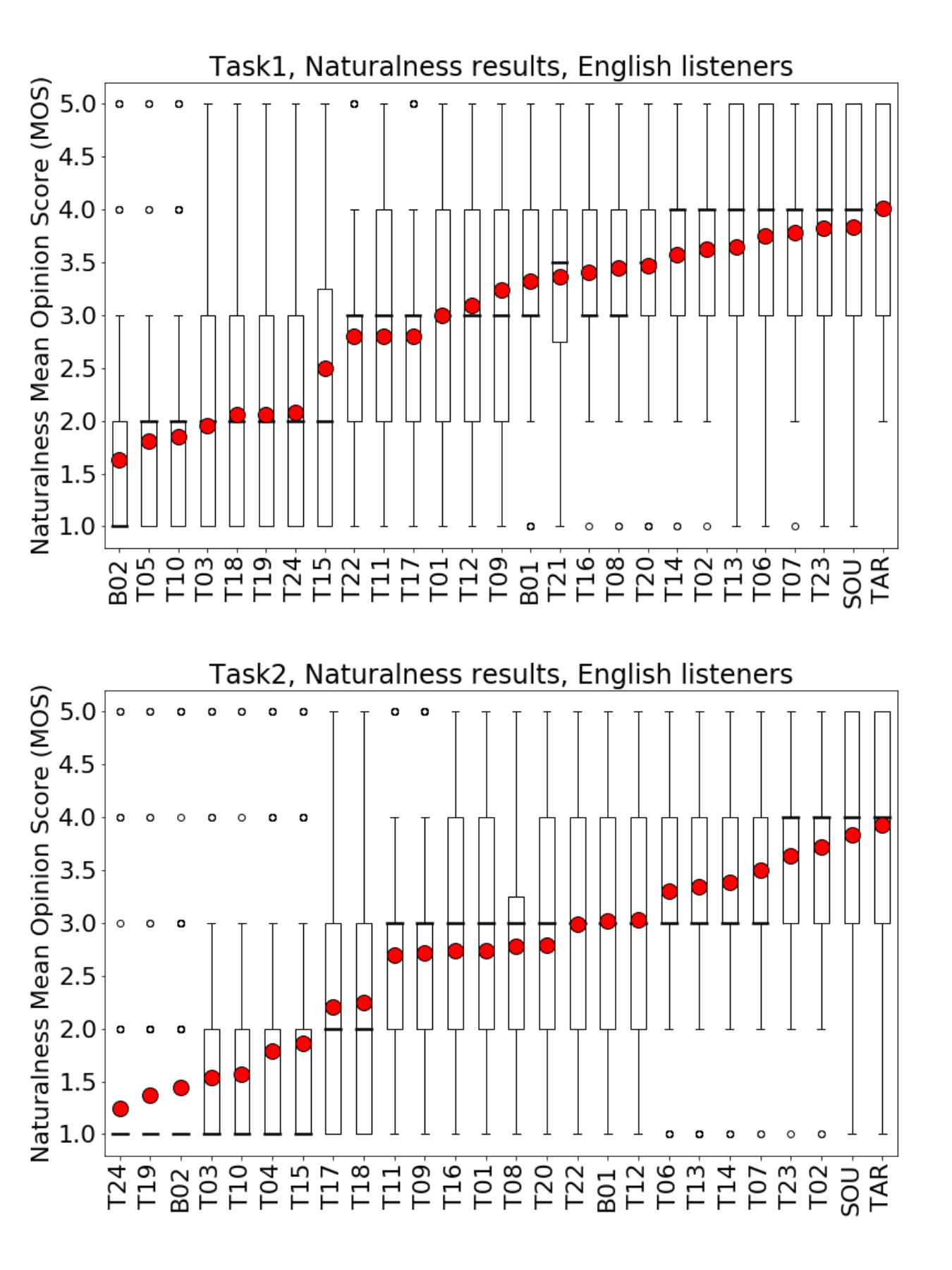}}
\end{minipage}
\begin{minipage}{0.4\linewidth}
       \subfigure[Japanese native speakers.]{
      \includegraphics[width=1\columnwidth]{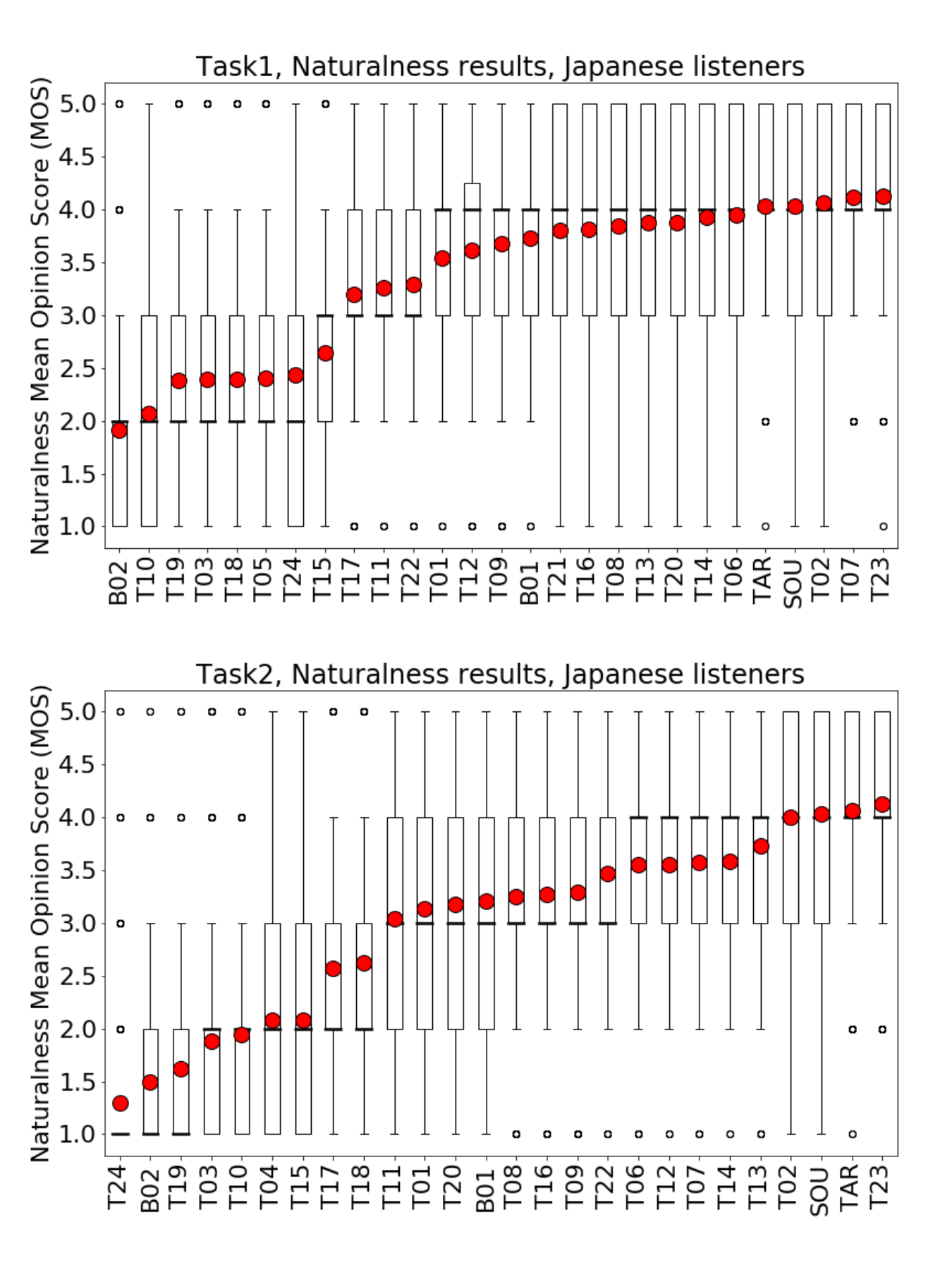}}
\end{minipage}

\caption{Official results on naturalness~\cite{huang2023singing}. The subjection evaluation is conducted by (a) English native speakers and (b) Japanese native speakers. The top column presents the results of Task1 (In-domain SVC), while the bottom shows that of Task2 (Cross-domain SVC). Our team number is T02.  }
\label{fig:3}
\end{figure*}

\begin{figure*}[htb]
  \centering
  \begin{minipage}{0.4\linewidth}
      \subfigure[English native speakers]{
      \includegraphics[width=1\columnwidth]{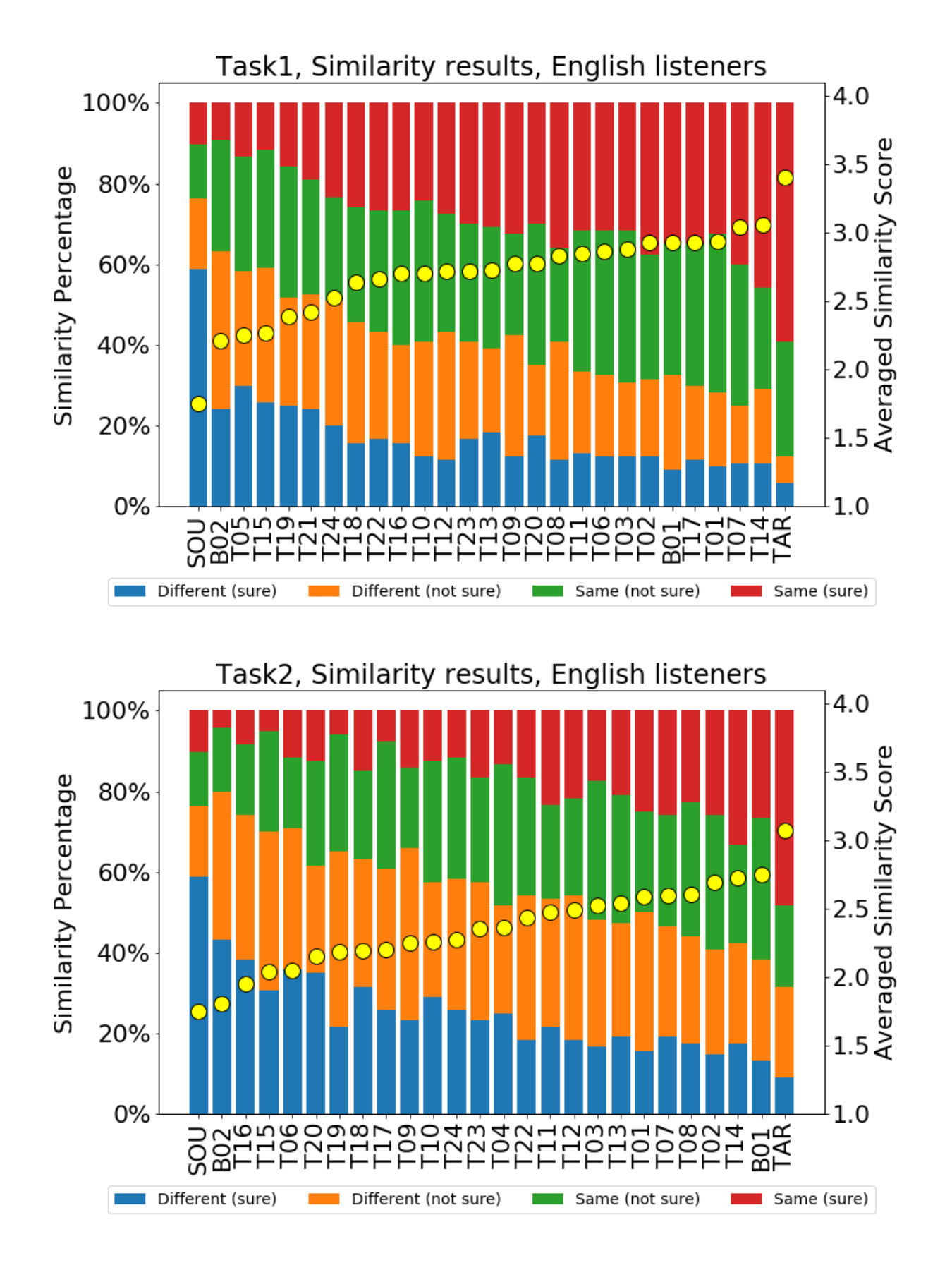}}
\end{minipage}
\begin{minipage}{0.4\linewidth}
       \subfigure[Japanese native speakers.]{
      \includegraphics[width=1\columnwidth]{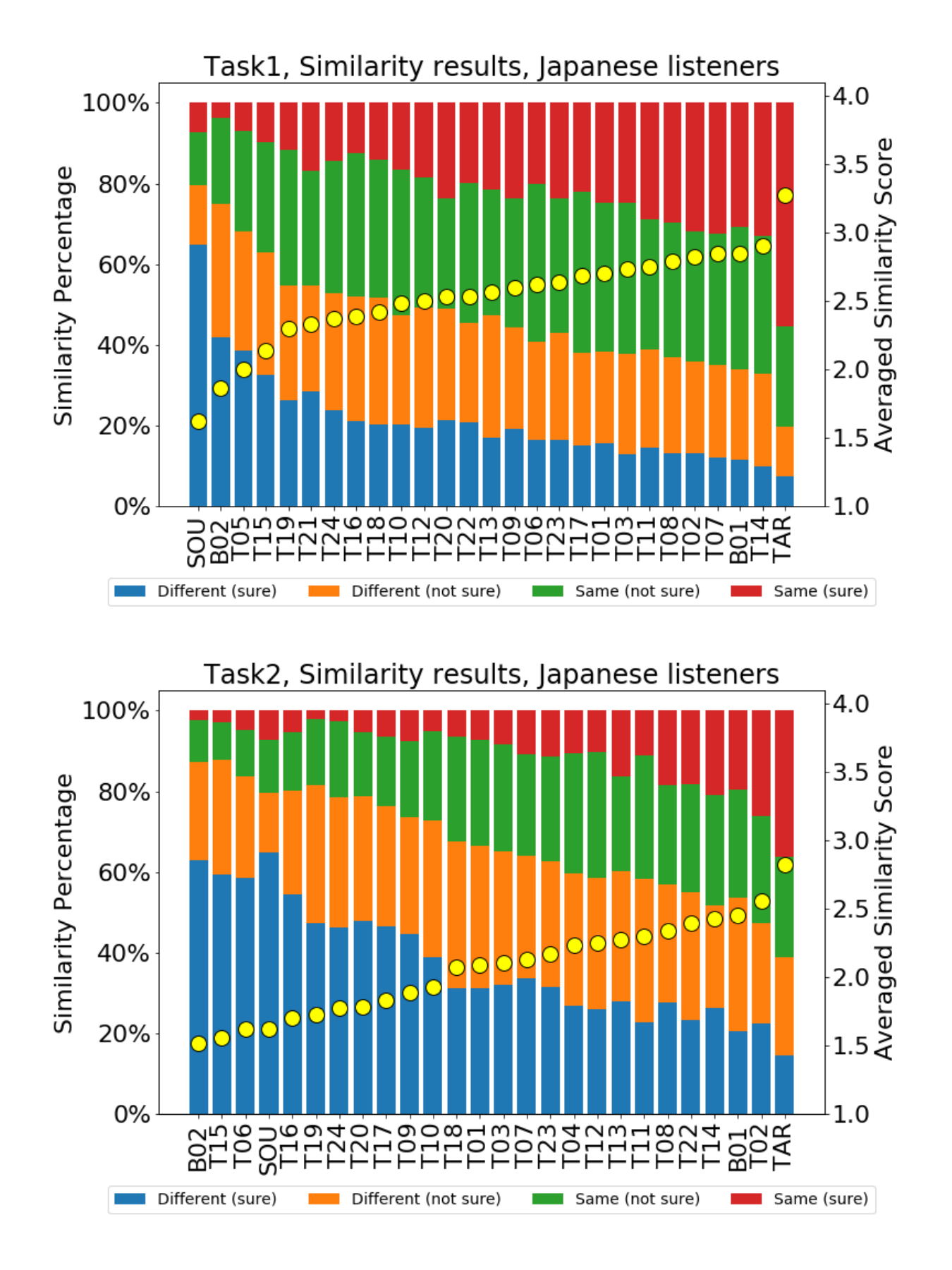}}
\end{minipage}
 
  \caption{Official results on similarity~\cite{huang2023singing}. The subjection evaluation is conducted by (a) English native speakers and (b) Japanese native speakers. The top column presents the results of Task1 (In-domain SVC), while the bottom shows that of Task2 (Cross-domain SVC). Our team number is T02. 
  }
  \label{fig:4}
  \end{figure*}

\section{EXPERIMENTS}
\label{sec:typestyle}
\subsection{Data and Model Training}
All involving datasets are listed in Table~\ref{tab:1}. For the voice converter, VCTK~\cite{yamagishi2019vctk} and a mixed singing dataset are involved in pre-training. The mixed singing dataset comprises about 73.3 hours selected from NUS48e~\cite{6694316}, Opencpop~\cite{wang2022opencpop}, M4singer~\cite{zhang2022msinger}, and Opensinger~\cite{huang2021multi}. In the in-domain SVC task, the singing data of two singers containing a total of 0.41 hours are provided for adaptation, while 0.18 hours of speech data are available in the cross-domain task. And the mixed speech dataset comprises 951.6 hours of speech recordings from LJspeech~\cite{ljspeech17}, VCTK~\cite{yamagishi2019vctk}, LibriTTS~\cite{zen2019libritts}, and HI-FI TTS~\cite{bakhturina2021hifi}.

For the voice converter, as the target singer's data in SVCC tasks is limited to roughly 10 minutes, we implement the standard pre-training and adaptation strategy. In the pre-training phase, we leverage the VCTK dataset to initiate our conversion model. Upon initiating the model with the VCTK, we then use the mixed singing dataset listed in Table 1 for additional pre-training. In the adaptation phase, we use the target speaker data and two auxiliary singers' data for fine-tuning. The two singers are the ones with the largest set of data selected from the mixed singing data. The auxiliary data helps in the stabilization of the adaptation process. 

We also incorporate speed perturbation~\cite{choi2021neural} based singing data augmentation in the adaptation phase. We adjust the speed of each audio clip from the target speaker by a random factor between 0.8 and 1.4, while maintaining the pitch, in order to enhance the diversity of rhythm. Specifically, we used the open-source Python library audiotsm\footnote{https://pypi.org/project/audiotsm/} to achieve the speed perturbation. Combining the original target speaker data, we get an augmented dataset that is twice as large as the original data. Using the augmented dataset for finetuning can effectively mitigate the SVC model's possible overfitting to the limited data.

The DSPGAN post-processor is first trained on the mixed speech dataset and then fine-tuned on the mixed singing dataset.

\subsection{Model Configuration}

For content extraction, the open-source SSL model\footnote{https://github.com/auspicious3000/contentvec}, a variant of HuBERT with further speaker-decoupling design, is used to extract 256-dim SSL features. As for the conversion model, we implement this model on an open-source SVC project\footnote{https://github.com/svc-develop-team/so-vits-svc}. Specifically, the posterior encoder utilizes six-layer non-causal WaveNet residual blocks, following WaveGlow~\cite{prenger2018waveglow}. The prior encoder is achieved by a multi-layer Transformer~\cite{vaswani2017attention}, which utilizes a six-layer Multi-Head Attention with two heads. The decoder follows the original configuration of the HIFI-GAN decoder in VITS. For the post-processer, we follow the parameterization and structure of DSPGAN~\cite{song2023dspgan}. During training, we train the conversion model for 600k and 300k steps in the mixed speech and singing dataset, respectively, with a batch size of 16. And 100k training steps are performed in the adaptation stage. With 1e-4 of the initiate learning rate, the conversion model is optimized with the Adam optimizer ($\beta_1$ = 0.8, $\beta_2$ = 0.99).

\subsection{Experimental Results for SVCC 2023}
Naturalness and similarity are evaluated in SVCC 2023. Naturalness refers to whether or not the synthesized vocals sound more human-like, with elements like electro-metal and unclear bites that can impact this aspect. Listeners are asked to evaluate the naturalness on a five-point scale. Similarity refers to the converted singing voice timbre resembling that of the target speaker's singing voice timbre. For assessing conversion similarity, a natural target speech and a converted speech are presented for listeners to judge whether the two samples are produced by the same speaker on a four-point scale. 
Subjective evaluation results for SVCC 2023 are presented in Fig.~\ref{fig:3} and Fig.~\ref{fig:4}. The results of the English listeners are considered as the main ones, while the results of the Japanese listeners are used only as a reference for evaluation. The original source and target speakers are referred to as \textit{SOU} and \textit{TAR}, respectively. The baseline model, \textit{B01}, follows the implementation of the DiffSVC~\cite{liu2021diffsvc}, while \textit{B02} represents the FastSVC~\cite{liu2021fastsvc}. All participating systems are indicated by the letter ``T" followed by a number and our system is identified as \textit{T02}.

\textbf{In-domain SVC: } The in-domain task aims to build an any-to-one SVC system using open-source singing datasets and the target speakers’ singing voices. According to the naturalness results in Fig.~\ref{fig:3}, our system T02 achieves the 5th for English listeners and the 3rd for Japanese listeners. According to the similarity results in Fig.~\ref{fig:4}, our system achieves the 5th for English listeners and the 3rd for Japanese listeners. These results show our system can achieve good naturalness and similarity in in-domain SVC task.

\textbf{Cross-domain SVC: }Different from the in-domain task, the cross-domain SVC task aims to synthesize natural singing voice of the target voice timbre with only speech data provided. According to the naturalness results in Fig.~\ref{fig:3}, our system T02 achieves the 1st for English listeners and the 2nd for Japanese listeners. According to the similarity results in Fig.~\ref{fig:4}, the similarity of our system achieves the 2nd for English listeners and the 1st for Japanese listeners. Compared to other teams, our system demonstrates superior naturalness and similarity in the cross-domain task.

\begin{table}[!h]
  \caption{Results of the ablation study.}
	\begin{tabular}{lll}
    \toprule
    Approach & Naturalness & Similarity \\
    \midrule
    Proposed Method  & $4.01 \pm 0.12 $&$ 3.43 \pm 0.15 $ \\
    \hspace*{0.2cm}w/o Speech Pre-training & $3.60 \pm 0.10 $&$ 3.21 \pm 0.09 $ \\
    \hspace*{0.2cm}w/o Adaptation Tricks& $3.90 \pm 0.07 $&$ 3.02 \pm 0.13 $\\
    \hspace*{0.2cm}w/o DSPGAN Post-processor & $3.52 \pm 0.08 $&$ 3.29 \pm 0.12 $ \\

    \bottomrule
  \end{tabular}
  \label{tab:2}
\end{table}

\subsection{Ablation study}
To further verify the effectiveness of our model, we conduct several ablations. Specifically, we remove the extra speech data pre-training and post-processing stage to evaluate its efficacy, forming the model variants \textit{w/o Speech Pre-training} and \textit{w/o Post-processing}. To evaluate the influence of our adaptation tricks, including data augmentation and auxiliary training strategy, we adapt the conversion model to the target speaker by directly fine-tuning using the data of the target speaker (\textit{w/o Adaptation Tricks}). To evaluate the performance of the ablations, the subjective evaluations of naturalness and similarity are both concocted. Twenty listeners are asked to participate in this evaluation. The results are presented in Table \ref{tab:2}. When removing the speech pre-training or post-processing stage, the naturalness gets degraded caused to the lower waveform reconstruction quality, while the similarity is also affected. Besides, without our specific-designed adaptation strategy, the system would easily overfit the limited training data, resulting in poor adaptation quality.

\section{Conclusions}
\label{sec:majhead}

This paper presents a VITS-based SVC system for the Singing Voice Conversion Challenge 2023, denoted as T02. The proposed system consists of a feature extractor, a voice converter, and a post-processor. The initial step entails the feature extractor processing linguistic contents from a singing signal. Then, a VITS-based conversion model generates the waveform of the target speaker. Furthermore, a DSPGAN vocoder is adopted to re-synthesize the converted waveform to remove the sound artifacts and improve the generation quality. As the data for the target singer is quite limited, we adopt a pretraining-adaptation schema, in which the conversion model is trained sequentially on speech and singing data and then adapted to the target singer. Besides, we implement various training tricks, such as data augmentation and joint training with auxiliary singer data, to alleviate the overfitting issue and get better adaptation performance. The offical evaluation results of SVCC 2023 demonstrate the superior performance of our system. Impressively, our system ranks the 1st and the 2nd in naturalness and similarity metrics, respectively, in the cross-domain task. Further ablations verify the effectiveness of our designs.

\section{Acknowledgements}
\label{sec:typestyle}

This work is supported in part by the National Key R\&D Program of China under Grant 2020AAA0109602.
\bibliographystyle{IEEEbib}
\bibliography{strings,refs}

\end{document}